\documentclass[conference, 9pt]{IEEEtran}
\IEEEoverridecommandlockouts
\usepackage{cite}
\usepackage{amsmath,amssymb,amsfonts}
\usepackage{algorithm}
\usepackage{algorithmic}
\usepackage{graphicx}
\usepackage{textcomp}
\usepackage{stfloats}
\usepackage{amsmath}
\usepackage{breqn}
\usepackage{stfloats}
\usepackage{multirow}
\usepackage{cite}
\usepackage{amsmath,amssymb,amsfonts}
\usepackage{adjustbox}
\usepackage{algorithm}
\usepackage{algorithmic}
\usepackage{graphicx}
\usepackage{textcomp}
\usepackage{stfloats}
\usepackage{amsmath}
\usepackage{breqn}
\usepackage{hyperref}
\usepackage{cite}
\usepackage{amsmath,amssymb,amsfonts}
\usepackage{algorithmic}
\usepackage{graphicx}
\usepackage{textcomp}
\usepackage{amsmath}
\ifCLASSOPTIONcompsoc
    \usepackage[caption=false, font=normalsize, labelfont=sf, textfont=sf]{subfig}
\else
\usepackage[caption=false, font=footnotesize]{subfig}
\fi
\usepackage{xcolor}

\def\BibTeX{{\rm B\kern-.05em{\sc i\kern-.025em b}\kern-.08em
    T\kern-.1667em\lower.7ex\hbox{E}\kern-.125emX}}
\begin{document}

\title{Mamba Meets Financial Markets: A Graph-Mamba Approach for Stock Price Prediction}

\author{\IEEEauthorblockN{Ali Mehrabian$^{1}$, Ehsan Hoseinzade$^{2}$, Mahdi Mazloum$^{3}$, and Xiaohong Chen$^{4}$}
\IEEEauthorblockA{$^{1}$Department of Electrical and Computer Engineering, The University of British Columbia, Vancouver, Canada \\
\hspace{-15pt}$^{2}$Department of Computer Science, Simon Fraser University, Burnaby, Canada \\
\hspace{-15pt}$^{3}$Department of Mathematics, University of Pittsburgh, Pittsburgh, USA \\
\hspace{-15pt}$^{4}$Department of Economics, Yale University, New Haven, USA\\
Email: alimehrabian619@ece.ubc.ca, ehoseinz@sfu.ca, sem322@pitt.edu, xiaohong.chen@yale.edu.
}
}

\maketitle
\begin{abstract}
Stock markets play an important role in the global economy, where accurate stock price predictions can lead to significant financial returns. While existing transformer-based models have outperformed long short-term memory (LSTM) networks and convolutional neural networks (CNNs) in financial time series prediction, their high computational complexity and memory requirements limit their practicality for real-time trading and long-sequence data processing.
To address these challenges, in this paper, we propose SAMBA, an innovative framework for stock return prediction that builds on the Mamba architecture and integrates graph neural networks (GNNs). SAMBA achieves near-linear computational complexity by utilizing a bidirectional Mamba block to capture long-term dependencies in historical price data and employing adaptive graph convolution to model dependencies between daily stock features. Our experimental results demonstrate that SAMBA significantly outperforms state-of-the-art baseline models in prediction performance, maintaining low computational complexity. The code and datasets are available at \href{https://github.com/Ali-Meh619/SAMBA}{github.com/Ali-Meh619/SAMBA}.
\end{abstract}

\section{Introduction}

Financial markets are one of the most crucial components of the global economy, with billions of dollars traded daily. Accurate predictions of stock market behavior can yield substantial financial gains for investors and institutions. However, the behavior of markets is inherently complex and difficult to forecast \cite{bb12,bb14}, as Fig. 1 depicts the fluctuating prices of the stock markets. This complexity has led industry and academia to focus on creating reliable models for predicting market movements, which are crucial for developing trading strategies, managing risk, and optimizing portfolios.

Recent advancements in deep learning have led to the application of various models for stock market prediction. Multi-layer perceptron (MLP)-based models capture complex, non-linear relationships in financial data \cite{chong2017deep,bb11}. Convolutional neural networks (CNNs), using 2D inputs of daily features or 3D inputs that include correlated stocks, extract temporal and inter-stock patterns \cite{hoseinzade2019cnnpred,hoseinzade2019u}. Long short-term memory (LSTM) networks are used to learn long-term dependencies in stock data, enhancing prediction performance \cite{bb7}.

The introduction of the self-attention mechanism \cite{vaswani2017attention} has further advanced stock market prediction by allowing models to capture global dependencies. Building on this, transformers have shown remarkable performance in sequence prediction tasks due to their ability to capture complex dependencies across different time steps, regardless of distance \cite{vaswani2017attention}. Transformers have been successfully applied to stock market prediction \cite{bb8,yoo2021accurate} by modeling intricate patterns in time series data; however, they come with drawbacks. Their quadratic complexity in inference can result in high computational costs and memory usage, limiting their practicality in real-time trading and scenarios involving long sequences.

To address these drawbacks, Mamba \cite{gu2023mamba} has been introduced as an efficient alternative for long-range sequence analysis. Mamba is based on the state space models (SSMs), which offer near-linear time complexity. Mamba incorporates a selective scan algorithm, which dynamically focuses on the most relevant portions of the input sequence while filtering out less useful data, thereby optimizing both computation and memory usage. A very recent MambaStock model applies a single Mamba model to predict stock prices using historical stock market data \cite{bb13}. However, a single Mamba model executes the selection mechanism in one direction only, which can limit its ability to capture global dependencies. To further capture the global dependencies, graph neural networks (GNNs) showed promising results in modeling correlations between stock features as a graph structure, capturing dependencies and co-movements in the market \cite{chen2018incorporating}. 

\begin{figure}[t]
    \includegraphics[width=0.48\textwidth]{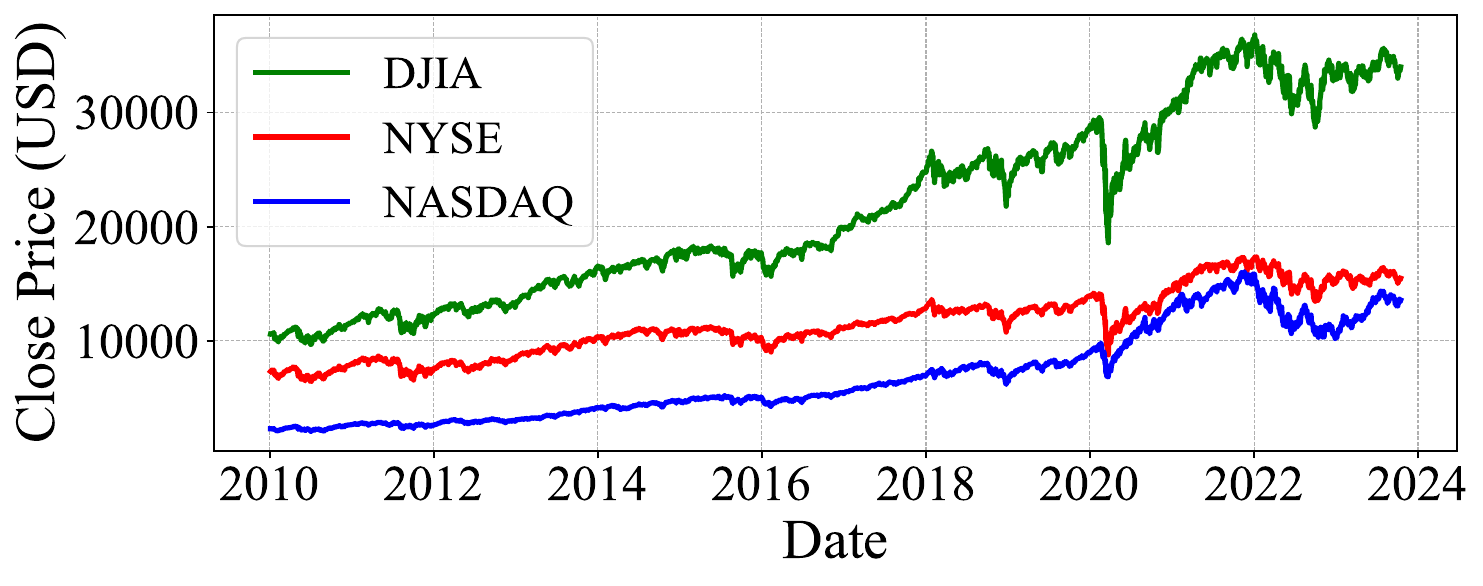}
    \vspace{-10pt}
    \caption{Close Price (in USD) for Dow Jones Industrial Average (DJIA), NASDAQ, and New York Stock Exchange (NYSE) stock markets from $2010$ to $2023$.}
    \label{fig:DJI}
    \vspace{-5pt}
\end{figure}


Building on the strengths of Mamba and GNNs, we introduce SAMBA, a novel model for stock market prediction that effectively handles complex sequential data. SAMBA consists of two main components: Bidirectional Mamba (BI-Mamba) block, which captures long-term dependencies in historical price data, and an adaptive graph convolutional (AGC) block, which models the interactions of daily stock features. Together, these components provide a comprehensive framework for accurate and efficient financial forecasting. Our key contributions are summarized as follows:

\begin{itemize} 
    \item \textbf{BI-Mamba for Long-Term Dependencies}: We incorporate BI-Mamba block to capture long-term dependencies within historical price data. This approach optimizes prediction performance and reduces computational complexity using the selective scan algorithm compared to transformers, making it suitable for real-time applications.
    \item \textbf{AGC block for Feature Interaction}: We employ an AGC block to capture and model the interactions between daily stock market features. In this framework, each daily feature is represented as a node, and the interactions between them are modeled as a graph structure. This enables SAMBA to effectively capture both temporal and relational patterns in the data, enhancing its predictive power.

    \item \textbf{Superior Performance Over State-of-the-art Baselines}: Extensive experiments demonstrate that SAMBA significantly outperforms several baseline algorithms in terms of predictive accuracy while maintaining low computational complexity, making it a robust tool for financial forecasting.

\end{itemize}

\section{Problem Formulation}
Following recent works on stock price prediction \cite{bb3,bb9,bb10}, our objective is to predict the change in stock price rather than the absolute value, framing this as a regression problem. Let $\mathcal{T}=\{1,2,\dots\}$ denote the set of days. The time interval $[t,t+1)$ is referred to as day $t\in \mathcal{T}$. Let $x_t^n$ denote the value for daily stock feature $n\in\mathcal{N}$ during day $t\in\mathcal{T}$. We denote the daily feature vector observed in day $t$ for all daily stock features as $\bold{x}_t=\left(x^1_t,\dots, x^N_t\right)\in \mathbb{R}^{N}$. We define the input daily feature matrix $\bold{X}=(\bold{x}_1,\dots,\bold{x}_L)^T\in \mathbb{R}^{L\times N}$, where $L$ is the number of historical observations from previous days. Daily stock features $\mathcal{N}$ include market-specific and general economic features (e.g., technical indicators, commodity prices, futures contracts) and they are obtained as explained in \cite{hoseinzade2019cnnpred}. Given the input daily feature data $\bold{X}$ from the past $L$ days, the objective is to predict the $1$-day return ratio as $o_{L+1} = \frac{c_{L+1}-c_L}{c_{L}}$, where $c_{L+1}$ and $c_L$ represent the closing stock prices at days $L+1\in \mathcal{T}$ and $L\in\mathcal{T}$, respectively.

\section{The Proposed SAMBA Model}
\subsection{Preliminaries on State Space Models}
Structured state space sequence models (S4) are a recent class of sequence models in deep learning that combine the characteristics of recurrent neural networks (RNNs) and CNNs \cite{b11}. They are inspired by SSMs used in control theory to describe the evolution of a continuous-time system's internal state. SSMs manage sequence-to-sequence transformations through an implicit latent state by using first-order differential equations. As an example, let \( \bold{x}(t) \in \mathbb{R}^L \) and \( \bold{y}(t) \in \mathbb{R}^L \) denote the continuous input and output functions, respectively. Let \( \bold{h}(t) \in \mathbb{R}^N \) denote the implicit state space. S4 models are characterized with four parameters \((\bold{\Delta}, \mathbf{A}, \mathbf{B}, \mathbf{C})\), which specify a function-to-function transformation through a continuous-time linear system SSM$(\bold{A},\bold{B},\bold{C})(\bold{x}(t))$ as follows:
\begin{equation}
\begin{aligned}
    \bold{h}'(t) = \mathbf{A}\bold{h}(t) + \mathbf{B}\bold{x}(t), \;\;\bold{y}(t) = \mathbf{C}\bold{h}(t),
\end{aligned}
\end{equation}
where \( \mathbf{A} \in \mathbb{R}^{N \times N} \), \( \mathbf{B} \in \mathbb{R}^{N \times L}\), and \( \mathbf{C} \in \mathbb{R}^{L \times N} \) are system parameters. The latent state compresses information about the past that can be accessed when processing the present input.
To perform the sequence-to-sequence transformation, the continuous parameters of the system can be discretized using a step size \(\bold{\Delta}\in \mathbb{R}^{N}\) by applying discretization methods such as the zero-order hold. Once discretized, the SSM can be represented as:
\begin{equation}
\begin{aligned}
    \bold{h}_k = \hat{\mathbf{A}}\bold{h}_{k-1} + \hat{\mathbf{B}}\bold{x}_k, \;\; \bold{y}_k = \mathbf{C}\bold{h}_k,
\end{aligned}
\end{equation}
where discretized matrices $\hat{\mathbf{A}}$ and $\hat{\mathbf{B}}$ are obtained as follows:
\begin{equation}\label{disc}
\begin{aligned}
\hat{\mathbf{A}} = \exp(\bold{\Delta} \mathbf{A}),\;\; \hat{\mathbf{B}} = (\bold{\Delta} \mathbf{A})^{-1} (\exp(\bold{\Delta} \mathbf{A}) - \mathbf{I}_N) \bold{\Delta} \mathbf{B}. 
\end{aligned}
\end{equation}
Transitioning from continuous \((\bold{\Delta}, \mathbf{A}, \mathbf{B}, \mathbf{C})\) to discrete form \(\hat{(\mathbf{A}}, \hat{\mathbf{B}}, \mathbf{C})\) allows for efficient computation using a linear recursive approach. The S4 model utilizes high-order polynomial projection operators (HiPPO) \cite{b12} to initialize the structure of the matrix $\bold{A}$, which improves long-range dependency modeling. Note that to operate over an input sequence $\bold{X}$ of length $L$ with $N$ daily stock features, the SSM is applied independently to each daily feature.
\vspace{-4pt}
\subsection{Bidirectional Mamba Block}
The Mamba model incorporates a data-dependent selection mechanism within the S4 framework and employs hardware-aware parallel algorithms. This selective scan algorithm is crucial for Mamba to effectively capture contextual information in long sequences while maintaining computational efficiency. By selectively focusing on the most relevant portions of the input data and filtering out irrelevant details, Mamba enhances its performance and reduces computational complexity for long sequence processing tasks.

\begin{algorithm}[t] \small
\caption{The process of Mamba model}
\begin{algorithmic}[1]
\STATE \textbf{Input:} \( \bold{X} \in \mathbb{R}^{L \times N} \).
\STATE \( \bold{X}_{\rm proj} \in \mathbb{R}^{L \times E} \gets \text{Projection}_E(\bold{X},b=0\)).
\STATE \( \bold{Z}_{\rm proj} \in \mathbb{R}^{L \times E} \gets \text{Projection}_E(\bold{X},b=0\)).
\STATE \( \bold{X}' \in \mathbb{R}^{L \times E} \gets \text{SiLU}(\bold{W}_{\rm conv}*\bold{X}_{\rm proj}) \).
\STATE \( \bold{B}_o \in \mathbb{R}^{L \times H} \gets \text{Projection}_H(\bold{X'},b=0) \).
\STATE \( \bold{C} \in \mathbb{R}^{L \times H} \gets \text{Projection}_H(\bold{X'},b=0) \).
\STATE \( \bold{A}_o \in \mathbb{R}^{E \times H} \gets \text{Learnable weight matrix}\).
\STATE \( \bold{\Delta} \in \mathbb{R}^{L \times E} \gets \text{Softplus}(\text{Projection}_E(\bold{X}',b=1)) \).
\STATE \( \hat{\bold{A}},\hat{\bold{B}} \in \mathbb{R}^{L \times E \times H} \gets \text{Discretize}(\bold{A}_o,\bold{B}_o,\bold{\Delta})\).
\STATE \( \bold{Y}_o \in \mathbb{R}^{L \times E} \gets \text{SSM}(\hat{\bold{A}},\hat{\bold{B}},\bold{C})(\bold{X}') \).
\STATE \( \bold{Y} \in \mathbb{R}^{L \times N} \gets \text{Projection}_N(\bold{Y}_o \odot \text{SiLU}(\bold{Z_{\rm proj}}),b=0) \).
\STATE \textbf{Return:} \( \bold{Y} \).
\end{algorithmic}\normalsize
\end{algorithm}
\setlength{\textfloatsep}{5pt}

We first define the linear projection operation for arbitrary input $\bold{Q}^{K \times N}$ as follows:
\begin{equation}
\bold{Q'}=\text{Projection}_{P}(\bold{Q},b=\{0,1\})=\bold{Q}\bold{W}+b\bold{b},
\end{equation}
where $\bold{W}\in \mathbb{R}^{N \times P}$ and $\bold{b}\in \mathbb{R}^{P}$ are learnable weights, and $b$ is a binary variable indicating the presence of a bias term. The Mamba model is summarized in Algorithm 1.
The process in the Mamba model begins with projecting the input matrix \( \bold{X} \in \mathbb{R}^{L \times N} \) into an embedding space of dimensionality \( E \) using a linear projection layer. This projection yields \( \bold{X}_{\rm proj} \in \mathbb{R}^{L \times E}\), which is further processed through a convolutional layer followed by a SiLU activation function to produce \( \bold{X}' \in \mathbb{R}^{L \times E}\). 
Simultaneously, two additional linear transformations generate matrices \( \bold{B}_o\in \mathbb{R}^{L \times H} \) and \( \bold{C}\in \mathbb{R}^{L \times H} \) with latent space dimension $H$ using matrix $\bold{X'}$, which are critical for parameterizing the SSM model. The key difference between classical S4 models and Mamba is that parameters $(\bold{B},\bold{C},\bold{\Delta})$ are functions of the input and for each time step $l\in\{1,\dots, L\}$, we have different state parameters. The state matrix \( \bold{A}_o\in \mathbb{R}^{E \times H} \) is also initialized as a learnable matrix. Then, the Softplus function is applied to the linear projection of  \( \bold{X}' \) to compute step size \( \bold{\Delta} \in \mathbb{R}^{L \times E} \).
Using $\bold{\Delta}$, discretized tensors \( \hat{\bold{A}} \in \mathbb{R}^{L \times E\times H}\) and \( \hat{\bold{B}} \in \mathbb{R}^{L \times E\times H}\) are obtained which define the dynamics of the SSM.
The SSM uses discretized tensors to process \( \bold{X}' \), capturing the dependencies in the data. The SSM model is time-varying based on the input and is applied at each time step $l\in\{1,\dots, L\}$ with a different set of state parameters. This provides the selective scan mechanism to selectively consider important information for each daily stock feature. As shown in Fig. 2(a), the output of the SSM is combined with $\bold{Z}_{\rm proj} \in \mathbb{R}^{L \times E}$ through a residual connection and is followed by a linear transformation to return the final output matrix \( \bold{Y} \in \mathbb{R}^{L \times N}\).

\begin{figure*}[!t]
    \centering
    \includegraphics[width=0.725\textwidth]{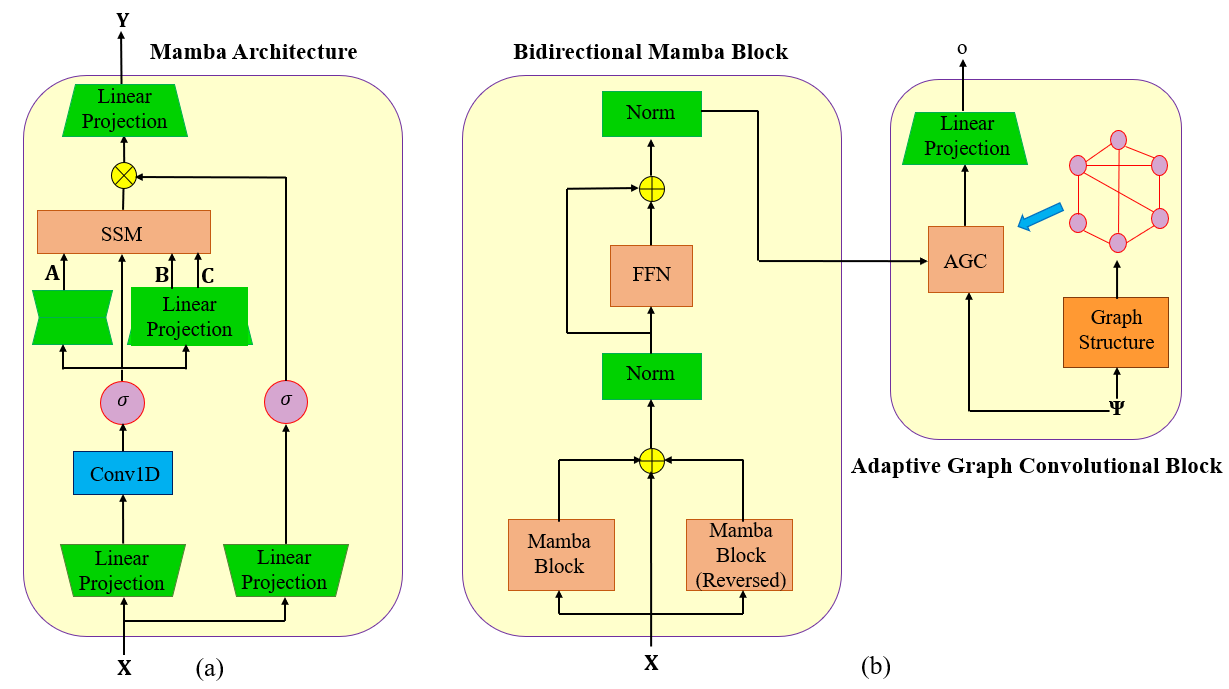}
    \vspace{0pt}
    \caption{Illustration of the proposed SAMAB model. (a) Architecture of the Mamba model. (b) The overall architecture of the bidirectional-Mamba block and adaptive graph convolutional block.}
    \label{fig:model}
    \vspace{-5pt}
\end{figure*}
We utilize the Mamba model to propose our BI-Mamba block to capture the temporal dependencies of daily stock features. As shown in \cite{b13}, a single Mamba model executes the selection mechanism in one direction only, which can limit its ability to capture global dependencies. To address this limitation, we propose BI-Mamba and AGC blocks to capture dependencies between daily features. The proposed BI-Mamba processes both the original and reverse input sequences in time to improve performance. As shown in Fig. 2(b), the input daily feature matrix is fed to Mamba models as follows:
\begin{equation}
    \bold{Y}_1=\text{Mamba}(\bold{X}),\;\; \bold{Y}_2=\text{Mamba}(\bold{P}\bold{X}),
\end{equation}
where $\bold{P}\in\ \mathbb{R}^{L \times L}$ is an anti-diagonal permutation matrix with the elements along the anti-diagonal being $1$, and all other elements being $0$. Through a residual connection, the input is added to outputs and normalized as follows:
\begin{equation}
    \bold{Y}_3=\text{Norm}(\bold{X}+\bold{Y}_1+\bold{P}\bold{Y}_2),
\end{equation}
where Norm is the layer normalization using mean and variance which can improve the convergence and training stability. The output $ \bold{Y}_3 \in \mathbb{R}^{L \times N}$ is fed through a feed-forward network (FFN) to capture the temporal dependencies and is normalized with a residual connection as follows:
\begin{align}
    \bold{Y}'&=\text{Projection}_L(\text{ReLU}(\text{Projection}_U(\bold{Y}_3^T,b=1)),b=1)^T,\nonumber\\
    \bold{Y}&=\text{Norm}(\bold{Y}'+\bold{Y}_3),
\end{align}
where $U$ is the hidden dimension. The proposed BI-Mamba block is repeated for $R$ layers, where $U$ and $R$ are hyperparameters, to capture the temporal dependencies of the processed daily stock features. Mamba uses a novel hardware-aware parallel computing algorithm to ensure efficient training through the backpropagation process.

\subsection{Adaptive Graph Convolutional Block}
To further capture the relationships between daily stock features, we model the interaction of daily stock features as a graph structure. The graph structure can characterize the interaction between the daily stock features in a fine-grained manner. We define an undirected weighted graph $\mathcal{G}(\mathcal{V},\mathcal{E})$ \cite{b16,b17}, where $\mathcal{V}=\{1,\dots,N\}$ is the set of nodes and $\mathcal{E}=\{(m,n)\,|\, m, n \in \mathcal{V}\}$ is the set of edges. Let $\bold{\tilde{A}}_{\rm \mathcal{G}}\in\mathbb{R}^{N\times N}$ denote the normalized adjacency matrix of graph $\mathcal{G}$, with each element $\bold{\tilde{A}}_{\rm \mathcal{G}}[m,n]$ corresponding to the normalized weight of the edge between nodes $m$ and $n$.

\begin{table*}[htbp]
\caption{\vspace{5pt}Comparison of prediction performance and computational efficiency between proposed SAMBA and the baseline methods. Best prediction performances are highlighted in \textbf{bold} and second-best prediction performances are \underline{underlined}.}
\vspace{-2pt}
\centering
\newcommand{\boldline}{\noalign{\hrule height 1.2pt}}
  \scalebox{0.87}{
\begin{tabular}{|c||c|c|c||c|c|c||c|c|c||c|c|c|}
\hline
\multirow{2}{*}{Method} & \multicolumn{3}{c||}{NASDAQ} & \multicolumn{3}{c||}{NYSE} & \multicolumn{3}{c||}{DJIA} & \multirow{2}{*}{\begin{tabular}[c]{@{}c@{}}Training\\ Time (sec/epoch)\end{tabular}} & \multirow{2}{*}{MACs} & \multirow{2}{*}{\begin{tabular}[c]{@{}c@{}}\# Parameters\end{tabular}} \\ \cline{2-10}
 & RMSE $\downarrow$ & IC $\uparrow$ & RIC $\uparrow$ & RMSE $\downarrow$ & IC $\uparrow$ & RIC $\uparrow$ & RMSE $\downarrow$ & IC $\uparrow$ & RIC $\uparrow$ &  &  &  \\ \boldline
LSTM \cite{bb7} & $0.0187$ & $0.0644$ & $0.0674$ & $0.0144$ & $0.0800$ & $0.0649$ & $0.0121$ & $0.0926$ & $0.0918$ & $0.081$ & $0.50$ M & $105,345$ \\ \hline
Transformer \cite{bb8} & $0.0147$ & $0.2440$ & $0.2542$ & $0.0141$ & $0.3189$ & $\underline{0.3293}$ & $0.0166$ & $0.2820$ & $\underline{0.3063}$ & $1.91$ & $13.77$ M & $402,658$ \\ \hline
FreTS \cite{bb4} & $\underline{0.0143}$ & $\underline{0.2722}$ & $\underline{0.2644}$ & $0.0143$ & $0.2324$ & $0.2208$ & $\underline{0.0116}$ & $0.2658$ & $0.2827$ & $0.514$ & $0.16$ M & $230,612$ \\ \hline
StockMixer \cite{bb3} & $0.0149$ & $0.2123$ & $0.1930$ & $0.0159$ & $0.2847$ & $0.2600$ & $0.0117$ & $0.2286$ & $0.2253$ & $0.169$ & $0.06$ M & $49,977$ \\ \hline
AGCRN \cite{b6} & $0.0147$ & $0.0948$ & $0.0932$ & $0.0142$ & $0.0757$ & $0.0772$ & $0.0169$ & $0.0607$ & $0.0742$ & $0.955$ & $61.09$ M & $149,304$ \\ \hline
FourierGNN \cite{bb5} & $0.0152$ & $0.1395$ & $0.1295$ & $0.0144$ & $0.1169$ & $0.1199$ & $0.0119$ & $0.1451$ & $0.1281$ & $1.10$ & $0.08$ M & $182,848$ \\ \hline
MambaStock \cite{bb13} & $0.0145$ & $0.2488$ & $0.2059$ & $\underline{0.0139}$ & $\underline{0.3697}$ & $0.3125$ & $0.0141$ & $\underline{0.3355}$ & $0.2776$ & $0.28$ & $0.06$ M & $85,524$ \\ \hline
SAMBA & $\textbf{0.0128}$ & $\textbf{0.5046}$ & $\textbf{0.4767}$ & $\textbf{0.0125}$ & $\textbf{0.5044}$ & $\textbf{0.4950}$ & $\textbf{0.0108}$ & $\textbf{0.4483}$ & $\textbf{0.4703}$ & $0.891$ & $0.11$ M & $167,178$ \\ \hline
\end{tabular}}
\end{table*}

Pre-determined graph structures, which rely on similarity measures or distance functions, are not directly tailored to specific prediction tasks, often leading to biases and poor system representations that can limit performance \cite{b5}. To address this issue, we propose a task-specific adaptive graph construction method that allows the model to learn dependencies between daily stock features as a graph structure in an end-to-end manner. We first initialize a learnable node embedding matrix $\bold{\Psi}\in \mathbb{R}^{N \times d_{\rm e}}$, where $d_{\rm e}$ is the node embedding dimension. We use the Gaussian kernel to build the dependency graph between daily stock features as follows:
\begin{equation}
\begin{aligned}
    \bold{D}&=\text{diag}(\bold{\Psi}\bold{\Psi}^T)\bold{1}_N^T+\bold{1}_N\text{diag}(\bold{\Psi}\bold{\Psi}^T)^T-2\bold{\Psi}\bold{\Psi}^T,\\
    \bold{\tilde{A}}_{\rm \mathcal{G}}&=\text{Softmax}(\text{exp}(-\psi\bold{D})),
\end{aligned}
\end{equation}
where $\text{diag}(\cdot)$ returns the diagonal elements of a matrix, $\bold{1}_N$ denotes an all-ones vector with dimension $N$, $\bold{D}\in \mathbb{R}^{N \times N}$ is the pairwise squared Euclidean distance matrix between the rows of the embedding matrix, and $\psi$ is the scaling factor which is initialized as a learnable scalar. The Softmax function is applied row-wise to normalize the obtained adjacency matrix. The Gaussian kernel can capture non-linear relationships between nodes that may not be apparent with linear measures like the dot product, leading to a more accurate and representative graph structure. Additionally, the \(\psi\) scaling factor controls how quickly the similarity between graph nodes decreases as their distance increases. A larger \( \psi \) makes the graph more sensitive to small differences, resulting in a more localized graph, while a smaller \( \psi \) leads to a broader, more connected graph. The node embedding matrix \(\bold{\Psi}\) is dynamically updated over time via the backpropagation algorithm during the learning process. As a result, the adjacency matrix of the daily stock features can be obtained in an end-to-end manner without requiring any prior knowledge.

Finally, we utilize graph convolutional operation \cite{b14} to perform graph spectral filtering with $K$-order approximation of the Chebyshev polynomial. The Chebyshev polynomial $T_n$ for arbitrary input $\bold{Q}^{N \times N}$ is defined as follows:
\begin{align*}
T_0(\mathbf{Q}) &= \mathbf{I}_N,\;\; T_1(\mathbf{Q}) = \mathbf{Q}, \\
T_n(\mathbf{Q}) &= 2\mathbf{Q}T_{n-1}(\mathbf{Q}) - T_{n-2}(\mathbf{Q}), \quad n \geq 2.
\end{align*}
We propose an AGC block that employs adaptive filter weights specific to each daily stock feature. We initialize the learnable filter weights \(\bold{W}_{\text{Filter}} \in \mathbb{R}^{N \times (K+1) \times L}\) and $\bold{b}_{\text{Filter}} \in \mathbb{R}^{N}$, providing a unique set of parameters for daily stock features. By sharing these parameters across all daily features, the model can dynamically learn the underlying patterns in each feature over time. This approach enables each feature to adapt and learn its specific patterns from a shared pool of parameters. The proposed AGC operation is expressed as follows:
\begin{equation}
\bold{o}'=\sum_{k=0}^KT_k( \bold{\tilde{A}}_{\mathcal{G}})\otimes\bold{W}_{\text{Filter}}[:,k+1,:]\otimes \bold{Y}+\bold{b}_{\text{Filter}},    
\end{equation}
where $\bold{o}'\in \mathbb{R}^{N}$ is the output of AGC block. The polynomial order $K$ is a hyperparameter that is fine-tuned in the training process. The final stock return output is generated by applying a linear transformation to the processed daily features as $o=\text{Projection}_1(\bold{o'}^T)\in \mathbb{R}$.

The number of daily features \( N \) can be large, which means that the weights \(\bold{W}_{\text{Filter}}\) and $\bold{b}_{\text{Filter}}$ would have a large number of parameters to train. This could lead to the overfitting problem, particularly if the dataset is not sufficiently large. To mitigate this issue, we employ a matrix factorization technique to reduce the number of parameters. Specifically, we use the embedding matrix \(\bold{\Psi}\) to generate the tensors \(\bold{W}_{\text{Filter}}\) and \(\bold{b}_{\text{Filter}}\). This is expressed as \(\bold{W}_{\text{Filter}} = \bold{\Psi} \otimes \bold{F}_{\rm w}\) and \(\bold{b}_{\text{Filter}} = \bold{\Psi}\bold{f}_{\rm b}\), where \(\bold{F}_{\rm w} \in \mathbb{R}^{d_{\rm e} \times (K+1) \times L}\) and \(\bold{f}_{\rm b} \in \mathbb{R}^{d_{\rm e}}\) are learnable weights. Since \(d_{\rm e} \ll N\), this approach significantly reduces the number of parameters that need to be trained, thus helping to prevent overfitting and improving the model's generalization.

\section{Performance Evaluation}
\textbf{Dataset and Implementation Details}: We apply our proposed model to three real-world datasets from the US stock market with $N=82$ daily stock features explained in \cite{hoseinzade2019cnnpred}: NASDAQ, New York Stock Exchange (NYSE), and Dow Jones Industrial Average (DJIA), covering the period from January 2010 to November 2023. We use the first $80\%$ of the data for training, the next $5\%$ for validation, and the rest $15\%$ for testing. We conduct simulations using a computing server with an Intel Silver $4216$ Cascade Lake @ $2.1$GHz CPU and four Nvidia Tesla V$100$ Volta GPUs with $32$ GB memory. We choose the parameters $E=64$, $H=64$, $R=3$, $U=32$, $K=3$, and $d_{\rm e}=10$ for the SAMBA model. To implement the neural networks, we use PyTorch library \cite{b87} and Adam optimizer \cite{b88}. The initial learning rate is set to $0.001$. We consider $1500$ epochs for training the models. The batch size is set to $128$. We choose $L=5$. We use the min-max normalization technique for training by scaling the data to be within the range of $[0,1]$.\\
\textbf{Baselines and Metrics}: We compare the performance of our proposed model with the following baselines: MLP-based models (i.e., StockMixer \cite{bb3}, FreTS \cite{bb4}), GNN-based models (i.e., AGCRN \cite{b6}, FourierGNN \cite{bb5}), Transformer \cite{bb8}, LSTM \cite{bb7}, and MambaStock \cite{bb13}. To evaluate the performance of the models, we employ three metrics. The first metric is the root mean squared error (RMSE). We consider two rank-based evaluation
metrics \cite{bb3,bb9}. The information coefficient (IC) shows how close the prediction is to the actual result, computed by the average Pearson correlation coefficient. The rank information coefficient (RIC) is based on the ranking of the stocks’ short-term profit potential, computed by the average Spearman coefficient. To measure the computational complexity, we consider training time per epoch and the total count of multiply-accumulate (MAC) operations. MAC operations involve multiplying two numbers and adding the result to an accumulator.\\
\textbf{Overall Comparison}: Table I presents the experimental results for the prediction models across three datasets and compares the computational efficiency. Our proposed SAMBA model outperforms all the baselines in RMSE, IC, and RIC metrics by a significant margin. In particular, SAMBA achieves improvements in RMSE, with gains of $11.72\%$ for NASDAQ, $10.07\%$ for NYSE, and $6.90\%$ for DJIA compared to the second-best models, respectively. The proposed SAMBA model also shows improvements in IC and RIC, providing $85.38\%$ and $80.30\%$ better performance for NASDAQ, $36.43\%$ and $50.32\%$ for NYSE, and $33.62\%$ and $53.54\%$ for DJIA compared to the second-best models, respectively. Additionally, the SAMBA model benefits from the BI-Mamba and AGC blocks, which improve performance compared to the MambaStock approach that relies on a one-directional Mamba model. Regarding efficiency, Table I shows that our proposed SAMBA model ranks fifth among all the models. It has a higher computational complexity than MLP-based models and MambaStock, but it provides a lower computational complexity than the GNN-based models and transformer architecture. In particular, the proposed SAMBA model improves the training time per epoch $6.70\%$ compared to the AGCRN model, and requires $0.11$ Million MACs, which is significantly less than that of the Transformer and AGCRN models.

\section{Conclusion}
The complex nature of stock markets makes predicting stock returns challenging. While transformer-based models have shown promising results, they require high computational resources. We introduce SAMBA, a computationally more efficient but still accurate solution for stock return prediction. SAMBA leverages the selective scan nature of the Mamba architecture to capture long-term patterns and uses adaptive graph convolution to infer the relationships between daily stock features. Our experiments show that SAMBA significantly outperforms several state-of-the-art baselines in prediction performance with low computational complexity, making it a valuable tool for real-time trading and financial analysis. For future work, we will investigate multi-step joint stock price prediction by considering correlations of stock markets.

\bibliographystyle{IEEEtran}
\footnotesize
\bibliography{ref}

\begin{thebibliography}{10}
\providecommand{\url}[1]{#1}
\csname url@samestyle\endcsname
\providecommand{\newblock}{\relax}
\providecommand{\bibinfo}[2]{#2}
\providecommand{\BIBentrySTDinterwordspacing}{\spaceskip=0pt\relax}
\providecommand{\BIBentryALTinterwordstretchfactor}{4}
\providecommand{\BIBentryALTinterwordspacing}{\spaceskip=\fontdimen2\font plus
\BIBentryALTinterwordstretchfactor\fontdimen3\font minus \fontdimen4\font\relax}
\providecommand{\BIBforeignlanguage}[2]{{%
\expandafter\ifx\csname l@#1\endcsname\relax
\typeout{** WARNING: IEEEtran.bst: No hyphenation pattern has been}%
\typeout{** loaded for the language `#1'. Using the pattern for}%
\typeout{** the default language instead.}%
\else
\language=\csname l@#1\endcsname
\fi
#2}}
\providecommand{\BIBdecl}{\relax}
\BIBdecl

\bibitem{bb12}
X.~Chen, J.~Racine, and N.~R. Swanson, ``Semiparametric {ARX} neural-network models with an application to forecasting inflation,'' \emph{IEEE Trans. Neural Netw.}, vol.~12, no.~4, pp. 674--683, 2001.

\bibitem{bb14}
H.~White, ``Economic prediction using neural networks: {T}he case of {IBM} daily stock returns,'' in \emph{Proc. IEEE Int. Conf. Neural Netw. (ICNN)}, San Diego, CA, Jul. 1988.

\bibitem{chong2017deep}
E.~Chong, C.~Han, and F.~C. Park, ``Deep learning networks for stock market analysis and prediction: Methodology, data representations, and case studies,'' \emph{Expert Syst. Appl.}, vol.~83, pp. 187--205, Apr. 2017.

\bibitem{bb11}
S.~Gu, B.~Kelly, and D.~Xiu, ``Empirical asset pricing via machine learning,'' \emph{Rev. Financ. Stud.}, vol.~33, no.~5, pp. 2223--2273, Feb. 2020.

\bibitem{hoseinzade2019cnnpred}
E.~Hoseinzade and S.~Haratizadeh, ``{CNN}pred: {CNN}-based stock market prediction using a diverse set of variables,'' \emph{Expert Syst. Appl.}, vol. 129, pp. 273--285, Mar. 2019.

\bibitem{hoseinzade2019u}
E.~Hoseinzade, S.~Haratizadeh, and A.~Khoeini, ``U-{CNN}pred: {A} universal {CNN}-based predictor for stock markets,'' \emph{arXiv preprint arXiv:1911.12540}, 2019.

\bibitem{bb7}
A.~Moghar and M.~Hamiche, ``Stock market prediction using {LSTM} recurrent neural network,'' \emph{Procedia Comput. Sci.}, vol. 170, pp. 1168--1173, Apr. 2020.

\bibitem{vaswani2017attention}
A.~Vaswani, N.~Shazeer, N.~Parmar, J.~Uszkoreit, L.~Jones, A.~N. Gomez, L.~u. Kaiser, and I.~Polosukhin, ``Attention is all you need,'' in \emph{Proc. Adv. Neural Inf. Process. Syst. (NIPS)}, Long Beach, CA, Dec. 2017.

\bibitem{bb8}
C.~Wang, Y.~Chen, S.~Zhang, and Q.~Zhang, ``Stock market index prediction using deep transformer model,'' \emph{Expert Syst. Appl.}, vol. 208, pp. 118--128, Jul. 2022.

\bibitem{yoo2021accurate}
J.~Yoo, Y.~Soun, Y.-c. Park, and U.~Kang, ``Accurate multivariate stock movement prediction via data-axis transformer with multi-level contexts,'' in \emph{Proc. ACM SIGKDD Int. Conf. Knowl. Discov. Data Min. (KDD)}, Virtual, Aug. 2021.

\bibitem{gu2023mamba}
A.~Gu and T.~Dao, ``Mamba: Linear-time sequence modeling with selective state spaces,'' \emph{arXiv preprint arXiv:2312.00752}, 2023.

\bibitem{bb13}
Z.~Shi, ``Mamba{S}tock: {S}elective state space model for stock prediction,'' \emph{arXiv preprint arXiv:2402.18959}, 2024.

\bibitem{chen2018incorporating}
D.~Cheng, F.~Yang, S.~Xiang, and J.~Liu, ``Financial time series forecasting with multi-modality graph neural network,'' \emph{Pattern Recognit.}, vol. 121, pp. 208--218, Aug. 2022.

\bibitem{bb3}
J.~Fan and Y.~Shen, ``{S}tock{M}ixer: {A} simple yet strong {MLP}-based architecture for stock price forecasting,'' in \emph{Proc. of AAAI Conf. on Artif. Intell.}, Vancouver, Canada, Feb. 2024.

\bibitem{bb9}
T.~Li, Z.~Liu, Y.~Shen, X.~Wang, H.~Chen, and S.~Huang, ``Master: {M}arket-guided stock transformer for stock price forecasting,'' in \emph{Proc. of AAAI Conf. on Artif. Intell.}, Vancouver, Canada, Feb. 2024.

\bibitem{bb10}
H.~Xia, H.~Ao, L.~Li, Y.~Liu, S.~Liu, G.~Ye, and H.~Chai, ``{CI-STHPAN}: {P}re-trained attention network for stock selection with channel-independent spatio-temporal hypergraph,'' in \emph{Proc. of AAAI Conf. on Artif. Intell.}, Vancouver, Canada, Feb. 2024.

\bibitem{b11}
A.~Gu, I.~Johnson, K.~Goel, K.~Saab, T.~Dao, A.~Rudra, and C.~R{\'e}, ``Combining recurrent, convolutional, and continuous-time models with linear state space layers,'' in \emph{Proc. Adv. Neural Inf. Process. Syst. \textit{(NeurIPS)}}, Virtual, Dec. 2021.

\bibitem{b12}
A.~Gu, T.~Dao, S.~Ermon, A.~Rudra, and C.~R{\'e}, ``Hi{PPO}: {R}ecurrent memory with optimal polynomial projections,'' in \emph{Proc. Adv. Neural Inf. Process. Syst. \textit{(NeurIPS)}}, Virtual, Dec. 2021.

\bibitem{b13}
Z.~Wang, F.~Kong, S.~Feng, M.~Wang, H.~Zhao, D.~Wang, and Y.~Zhang, ``Is {M}amba effective for time series forecasting?'' \emph{arXiv preprint arXiv:2403.11144}, 2024.

\bibitem{b16}
A.~Mehrabian and V.~W.~S. Wong, ``Joint spectrum, precoding, and phase shifts design for {RIS}-aided multiuser {MIMO} {THz} systems,'' \emph{IEEE Trans. Commun.}, vol.~72, no.~8, pp. 5087--5101, Aug. 2024.

\bibitem{b17}
------, ``Adaptive bandwidth allocation in multiuser {MIMO} {TH}z systems with graph-transformer networks,'' in \emph{Proc. of IEEE Int. Conf. Commun. (ICC)}, Denver, CO, Jun. 2024.

\bibitem{bb4}
K.~Yi, Q.~Zhang, W.~Fan, S.~Wang, P.~Wang, H.~He, N.~An, D.~Lian, L.~Cao, and Z.~Niu, ``Frequency-domain {MLP}s are more effective learners in time series forecasting,'' in \emph{Proc. Adv. Neural Inf. Process. Syst. \textit{(NeurIPS)}}, New Orleans, LA, Dec. 2023.

\bibitem{b6}
L.~Bai, L.~Yao, C.~Li, X.~Wang, and C.~Wang, ``Adaptive graph convolutional recurrent network for traffic forecasting,'' in \emph{Proc. Adv. Neural Inf. Process. Syst. (NeurIPS)}, Virtual, Dec. 2020.

\bibitem{bb5}
K.~Yi, Q.~Zhang, W.~Fan, H.~He, L.~Hu, P.~Wang, N.~An, L.~Cao, and Z.~Niu, ``{F}ourier{GNN}: {R}ethinking multivariate time series forecasting from a pure graph perspective,'' in \emph{Proc. Adv. Neural Inf. Process. Syst. \textit{(NeurIPS)}}, New Orleans, LA, Dec. 2023.

\bibitem{b5}
A.~Mehrabian, S.~Bahrami, and V.~W. Wong, ``A dynamic {B}ernstein graph recurrent network for wireless cellular traffic prediction,'' in \emph{Proc. IEEE Int. Conf. Commun. (ICC)}, Rome, Italy, May 2023.

\bibitem{b14}
T.~N. Kipf and M.~Welling, ``Semi-supervised classification with graph convolutional networks,'' in \emph{Proc. Int. Conf. on Learning Representations (ICLR)}, Toulon, France, Apr. 2017.

\bibitem{b87}
A.~Paszke, S.~Gross, F.~Massa, A.~Lerer, J.~Bradbury, G.~Chanan, T.~Killeen, Z.~Lin, N.~Gimelshein, L.~Antiga \emph{et~al.}, ``Py{T}orch: An imperative style, high-performance deep learning library,'' in \emph{Proc. Adv. Neural Inf. Process. Syst. (NeurIPS)}, Vancouver, Canada, Dec. 2019.

\bibitem{b88}
D.~P. Kingma and J.~Ba, ``Adam: {A} method for stochastic optimization,'' in \emph{Proc. Int'l Conf. Learn. Representations ({ICLR})}, San Diego, CA, May 2015.

\end{thebibliography}

\end{document}